\documentclass[aip,
 amsmath,amssymb,
 reprint,%
]{revtex4-1}
\usepackage{graphicx}% Include figure files
\usepackage{dcolumn}% Align table columns on decimal point
\usepackage{bm}% bold math
\usepackage[percent]{overpic}
\usepackage{xcolor}
\usepackage[utf8]{inputenc}
\usepackage[T1]{fontenc}
\usepackage{mathptmx}
\usepackage{etoolbox}
\usepackage{algorithm}
\usepackage{algpseudocode}
\usepackage{hyperref}
\usepackage{tikz}
\usepackage{ulem} % for \sout{} command
\usetikzlibrary{arrows.meta}
\usetikzlibrary{angles,quotes}
\usetikzlibrary{arrows.meta, positioning}
\makeatletter
\def\@email#1#2{%
 \endgroup
 \patchcmd{\titleblock@produce}
  {\frontmatter@RRAPformat}
  {\frontmatter@RRAPformat{\produce@RRAP{*#1\href{mailto:#2}{#2}}}\frontmatter@RRAPformat}
  {}{}
}%
\makeatother
\begin{document}
\preprint{AIP/123-QED}

\title{FIREWALL: A surrogate model for the rapid assessment of tokamak wall loading and melting by runaway electrons}
% Force line breaks with \\
\author{V.J. Svensson}
\affiliation{ 
Max Planck Institute for Plasma Physics, Boltzmannstr. 2, 85748, Garching, Germany}
\author{T. Rizzi}
\affiliation{
Department of Electromagnetics and Plasma Physics, KTH Royal Institute of Technology, 100 44, Stockholm, Sweden}
\author{S. Ratynskaia}
\affiliation{
Department of Electromagnetics and Plasma Physics, KTH Royal Institute of Technology, 100 44, Stockholm, Sweden}
\author{H. Bergström}
\affiliation{ 
Max Planck Institute for Plasma Physics, Boltzmannstr. 2, 85748, Garching, Germany}
\author{L.V. Greco}
\affiliation{ 
Max Planck Institute for Plasma Physics, Boltzmannstr. 2, 85748, Garching, Germany}
\author{M. Hoelzl}
\affiliation{ 
Max Planck Institute for Plasma Physics, Boltzmannstr. 2, 85748, Garching, Germany}
\affiliation{
Department of Physics and Astronomy, Chalmers University of Technology, Göteborg, 41296, Sweden}
\author{P. Tolias}
\affiliation{
Department of Electromagnetics and Plasma Physics, KTH Royal Institute of Technology, 100 44, Stockholm, Sweden}
\date{\today}% It is always \today, today,
\begin{abstract}

Runaway electron (RE) beams generated during tokamak disruptions can deposit highly localized heat loads on plasma-facing components, posing a serious risk of melting and damage. Monte Carlo particle transport simulations coupled with three-dimensional thermomechanical response modeling can quantify this damage but are too computationally demanding for device-scale assessments and extensive scenario scans. We present FIREWALL (Fast Integrated Runaway Electron WALL loads), a surrogate model that combines a database of \textsc{Geant4} volumetric energy-deposition profiles with a one-dimensional heat-diffusion solver for each wall element. FIREWALL retains the energy and incident angle distributions of impacting REs and predicts the spatiotemporal temperature evolution of detailed three-dimensional wall geometries up to the melting threshold.  FIREWALL thus provides a fast physics-based framework for translating global RE simulations into global wall melting predictions, enabling large-scale screening of disruption scenarios while directing high-fidelity costly workflows to the limited wall regions where they are actually required.
\end{abstract}
\maketitle

\section{Introduction}\label{sec1}

In tokamaks, the runaway electrons (REs) generated during disruptions~\cite{Breizman2019} can pose a major threat to the heat handling capabilities and mechanical integrity of plasma-facing components (PFCs) and underlying structures~\cite{PITTS, Ratynskaia2}. REs reach relativistic energies and can be deposited over narrow surface areas, resulting in highly localized heat loads that may cause deep volumetric heating, severe melting, intense vaporization, or explosive damage to PFCs, thereby threatening machine integrity and even cooling or magnetic infrastructure. Ongoing efforts in the community aim to minimize the formation of massive RE beams in ITER-scale devices~\cite{re-formation, VottaAccepted} or to achieve more benign RE beam termination with reduced PFC heat loads~\cite{Reux_PRL, Paz-Soldan2021, Sheikh_2024, benign_hoppe}. In spite of these efforts, the complete elimination of the RE risk is not presently feasible, making the accurate prediction of the wall response to RE impacts essential for optimizing engineering designs and preparing safe experimental activities in ITER-like devices before entering high-current operation. As a consequence, extensive experimental and theoretical research is being pursued to understand and quantify RE-induced PFC damage~\cite{Ratynskaia2}. The present work addresses an important practical gap by introducing a surrogate model for the rapid assessment of material response to RE impacts. 

Accurate estimates of PFC damage first require predictions of the spatiotemporal wall distribution of RE impacts, which is already challenging, as it entails modeling RE formation and dynamics in a disrupting plasma, where quantities such as density, temperature, impurity content, magnetic-field topology and electric field vary drastically in space and time~\cite{Ratynskaia2, Bergstrom2024, Vannini_2025}. Tools of varying levels of complexity have been developed for this purpose, ranging from fast lower-dimensional models~\cite{dream} to self-consistent three-dimensional nonlinear MHD codes~\cite{Hoelzl_2021, Bandaru_2024, RE_fluid, Bergstrom2025, M3d-c1, M3d-c1_re}. To obtain a complete picture of RE-induced wall heating and damage, such modeling must provide the spatiotemporally varying impact angle and incident energy distributions of REs on wall elements. In high-fidelity wall damage modeling, this information serves as input to Monte Carlo (MC) simulations of the energy deposition, which in turn provide the volumetric source for coupled thermomechanical~\cite{Ratynskaia_graphite, Ratynskaia2, Rizzi_graphite} or heat transfer simulations~\cite{Ratynskaia1, Ratynskaia2, Rizzi_SPARC}. The MC simulations account for the full wall panel geometry and magnetic field topology, together with the spatially dependent RE wetting, to capture the energy losses and, in particular, the RE and cascade particle backscattering, transmission and re-deposition processes~\cite{Ratynskaia_graphite, Ratynskaia1, Ratynskaia2}. The combination of large tangential scales with the very high in-depth resolution required for shallow incidence angles makes such MC simulations computationally demanding. Heat transfer simulations likewise require high in-depth resolution to resolve the steep gradients of the volumetric source and the erosion of the free surface due to vaporization \cite{Ratynskaia1, Ratynskaia2}.

Reduced modeling tools, like RACLETTE~\cite{Raclette_1,Raclette_2} \& HEAT~\cite{HEAT_1}, have been successfully employed in fusion research for a quick assessment of heat loading and thermal responses~\cite{Raclette_3,Raclette_4,HEAT_2}. The REs pose a distinct challenge, since relativistic electrons deposit their energy volumetrically with the deposition profile depending strongly on their incident energy and impact angle. These features of RE–wall interactions must therefore be captured in a reduced description. Here we propose FIREWALL (Fast Integrated Runaway Electron WALL loads), a surrogate model for rapid wall melting assessment that retains the essential physics of RE energy deposition. The model reconstructs the volumetric energy-deposition profile from an MC simulation database and subsequently solves a one-dimensional heat diffusion equation for the temperature evolution within each wall element. Phase change as well as cooling and mass loss due to vaporization are neglected. Hence, owing to the short, typically millisecond, loading times, the predictions are reliable up to the onset of melting. FIREWALL, thus, enables a rapid assessment of the threat posed by incoming RE beams by providing the in-vessel temperature distribution across the entire device up to the melting threshold. For instance, the global analysis of a complete ITER RE beam termination simulation takes only a few minutes on a typical 256-core computing node, enabling a large number of disruption scenarios to be scanned in support of future reactor design studies.

The paper is organized as follows. Sec.~\ref{sec2} introduces the reduced model and its numerical implementation in FIREWALL. Sec.~\ref{sec3} features its verification against the high-fidelity \textsc{Geant4}--MEMENTO workflow~\cite{Ratynskaia1}. Sec.~\ref{sec4} presents a production-scale application to a full-domain ITER simulation carried out with JOREK. Finally, Sec.~\ref{sec5} summarizes the main conclusions and outlines future work.

\section{Model implementation}\label{sec2}

\begin{figure*}[!t]
    \centering
        \includegraphics[width=0.269\linewidth]{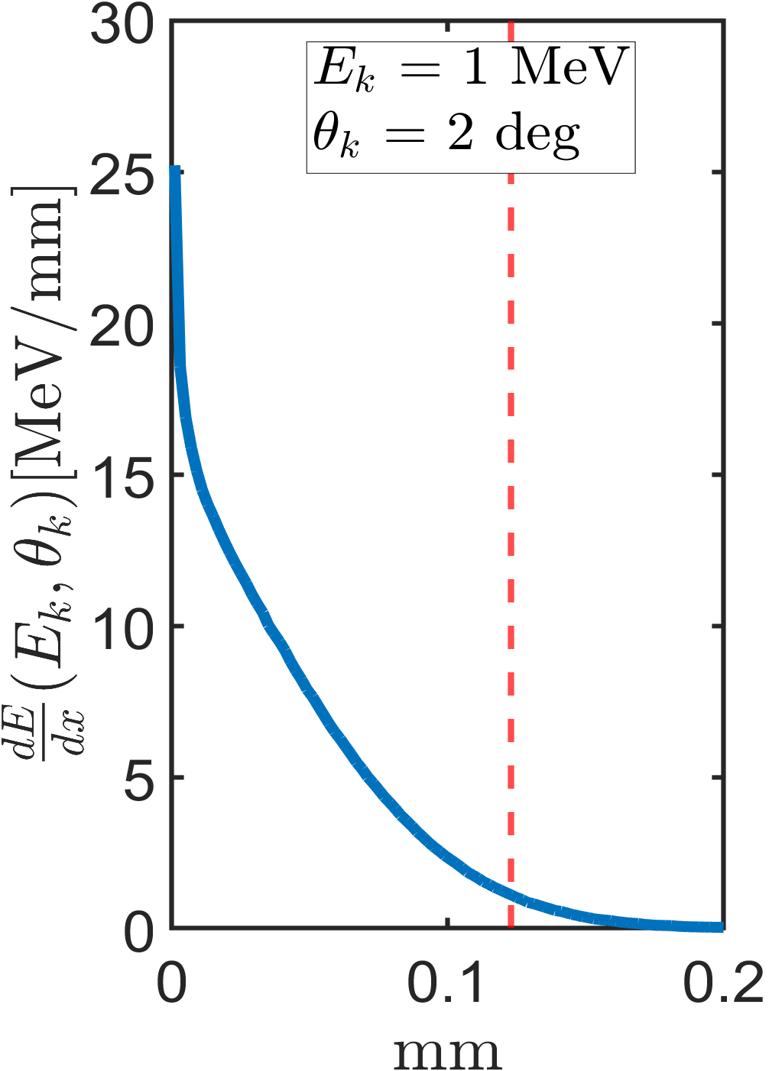}%
        \label{fig:Example_1}
    \hspace{0.001cm}
        \includegraphics[width=0.259\linewidth]{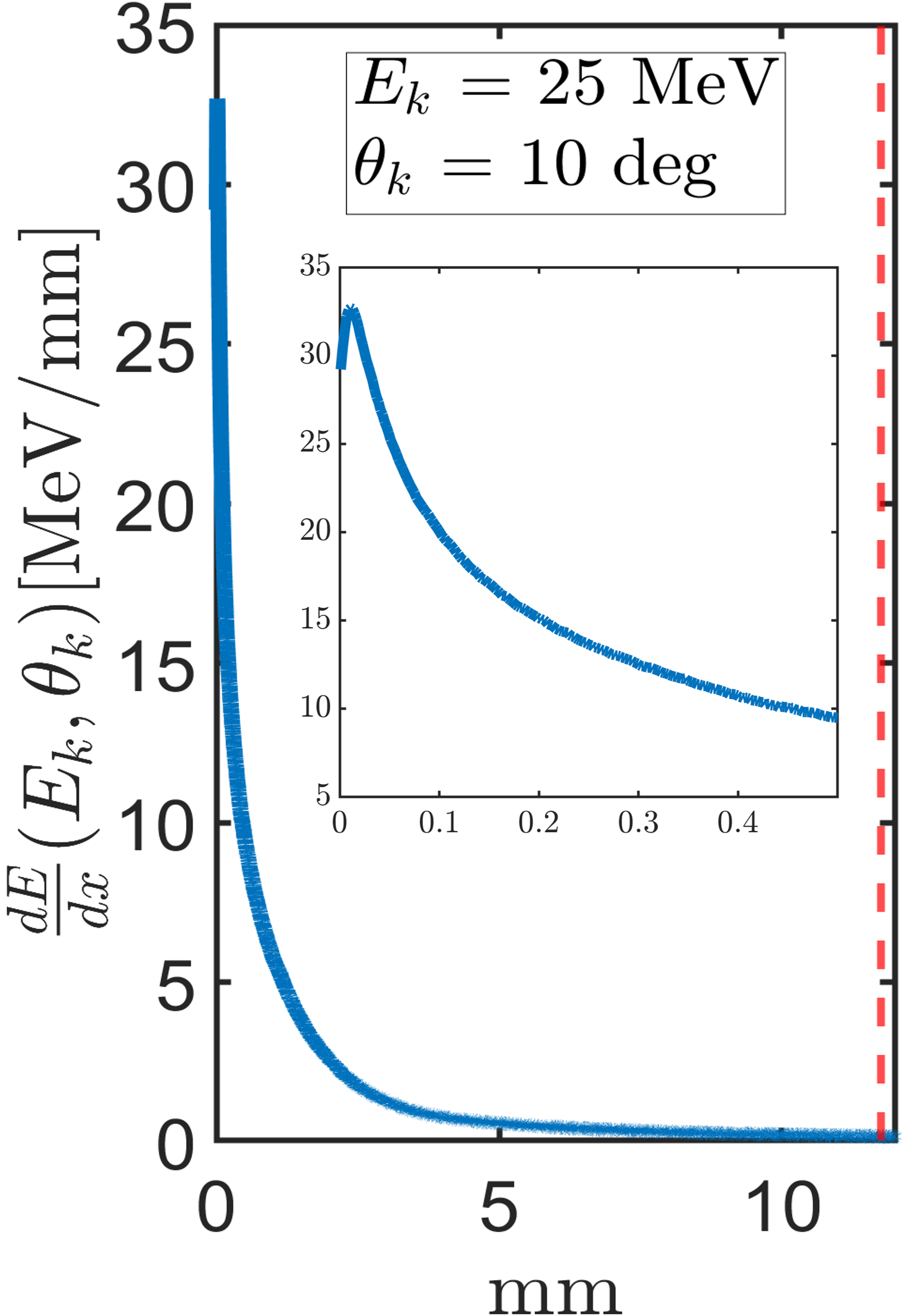}%
        \label{fig:Example_2}
    \hspace{0.01cm}
        \includegraphics[width=0.41\linewidth]{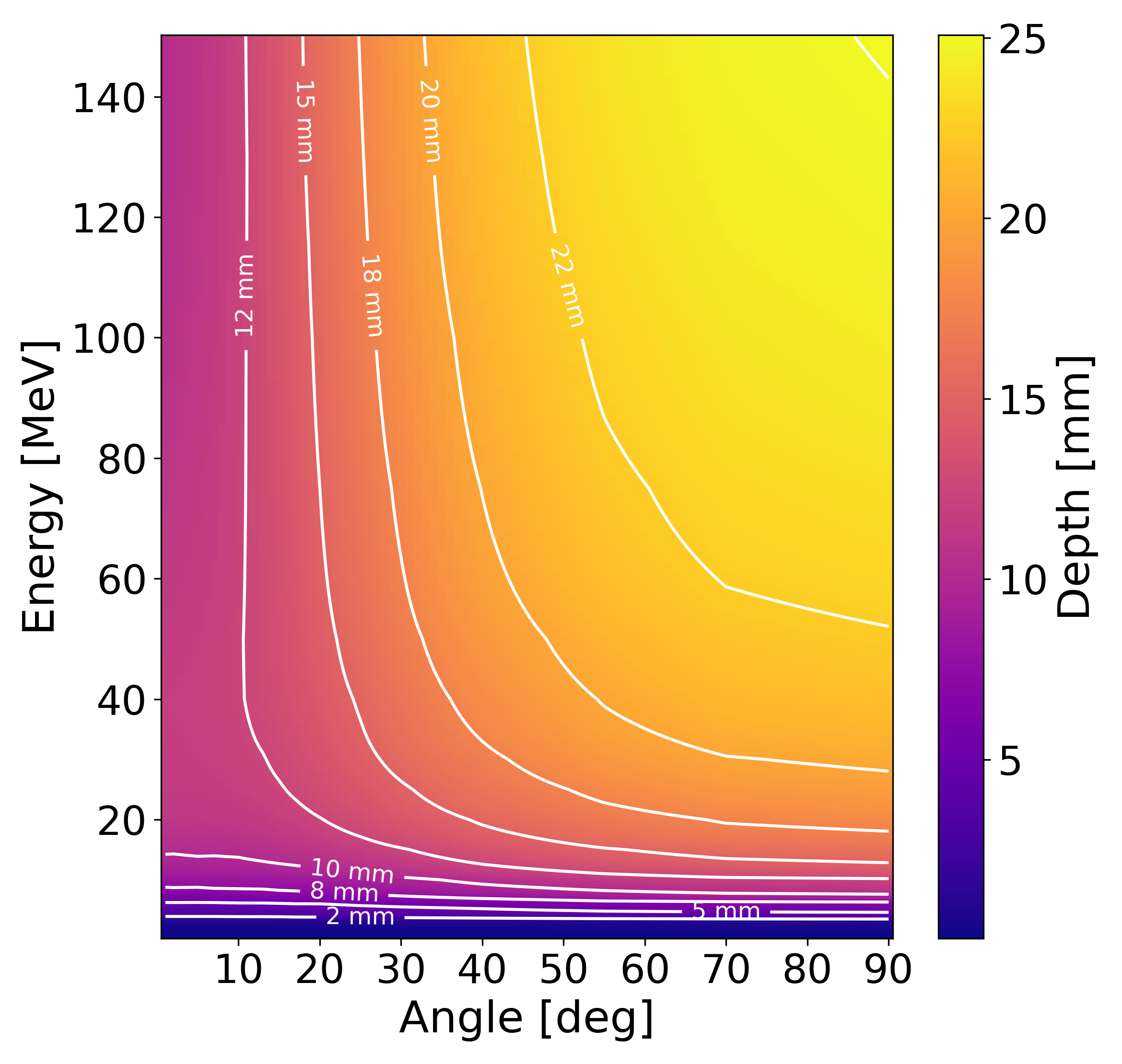}%
        \label{fig:dataset}
    \caption{(a),(b) Examples of in-depth energy deposition profiles $dE/dx(E_k,\theta_k,x)$ for two combinations of initial $(E_k,\theta_k)$, as specified in the inserts. The total deposited energy, $\int_0^{x_d} dE/dx(E_k,\theta_k) dx$, of 0.92\,MeV for (a) and 22.11\,MeV for (b), signifies energy losses through escaping products. (c) The depth corresponding to $95\%$ of the deposited energy for different $(E_k,\theta_k)$ combinations from the interpolation of the whole \textsc{Geant4} dataset. This depth is also indicated with a red dashed line in panels (a) and (b). }
    \label{fig:interpolation_data}
\end{figure*}

\subsection{Energy deposition}

At the core of the FIREWALL calculations lies the surrogate model for the RE energy deposition. Namely, a database of \textsc{Geant4}\,\cite{Allison_2016} MC simulations has been generated to enable the reconstruction of one-dimensional energy deposition profiles within a wall element.

Without loss of generality, the simulations were performed in a 3D rectangular domain with tangential dimensions of $100\,\mathrm{mm}\times200\,\mathrm{mm}$ and a thickness of $24\,\mathrm{mm}$. Since here we are concerned with 1D profiles, only in-depth meshing with a 2$\,\mu$m resolution is built. For more detailed discussion on the depth refinement see Refs.\,\cite{Ratynskaia1,Ratynskaia2}. These dimensions were selected to approximate a ``semi-infinite wall'' configuration: the depth captures the continuous slowing down approximation (CSDA) range of the highest-energy electron considered~\cite{ESTAR}, while the tangential dimensions extend over several Larmor radii and multiple helical pitches in the directions perpendicular and parallel to the magnetic field. This ensures that possible re-deposition of the back-scattered electrons is properly accounted for.  In \textsc{Geant4}, a point-like particle source is implemented that samples different combinations of the initial kinetic energy $E_k$ and of the incident angle $\theta_k$. Each $(E_k,\theta_k)$ pair is simulated with Monte Carlo statistics of $10^6$ primary particles striking the center of the top surface. Such a set-up is equivalent to uniform loading, as opposed to high-fidelity Geant4 simulations, where all tangential variations in the loading are preserved on the actual 3D panel geometry \cite{Ratynskaia1}.

To ensure the applicability of the model to various fusion devices, the dataset is being expanded to different wall materials, magnetic field strengths and magnetic field inclination angles. Meanwhile, in this work we demonstrate the tool capabilities and its accuracy focusing on recently studied ITER relevant scenarios\cite{Bergstrom2025, Bandaru_2024, Ratynskaia1}. 

Here we present results from the MC database for a full tungsten wall and a magnetic field of $7~\mathrm{T}$ that is inclined by $5^\circ$ with respect to the surface. The explored parameter space spans $E_{k}=[0.5,1.0,2,3,6,10,15,20,25,30,40,50,75,100,115,130,150]$ MeV and $\theta_k = [1,2,5,7,10,13,15,20,25,30,40,55,70,90]$ degrees, with $\theta_k$ defined with respect to the surface tangent. For each $(E_k,\theta_k)$ combination, a 1D in-depth energy deposition profile $dE/dx(E_k,\theta_k;x)$ is obtained. It is noted that the deposited energy takes into account losses from the escape of secondary particles \cite{Ratynskaia1, Ratynskaia2}.

The calculations are based on the selection of a computationally efficient and accurate \textsc{Geant4} physics list that covers all electromagnetic processes governing the transport of primary REs as well as the generation and subsequent transport of secondary particles until their instantaneous CSDA range is much smaller than the mesh size~\cite{Ratynskaia1}. The simulations employ a detailed physical description of relativistic electron interactions with matter through the \texttt{G4EmStandardSS} electromagnetic physics library, which relies on single-scattering models for the interaction of charged particle with the dressed nuclei, based on Dirac partial wave analysis with the \texttt{ELSEPA} code\,\cite{Salvat_2021} and incorporates PENELOPE\,\cite{Penelope} and Livermore\,\cite{Depaola_2006,Depaola_2003,Depaola_2000} based models for all other electromagnetic interactions. The simulations also include all relevant nuclear processes that dictate RE-induced neutron release and transport. This aspect of the database barely affects energy deposition, but is important for activation studies, thus, it will be discussed elsewhere. 

Examples of results for such MC simulations are presented in Figure~\ref{fig:interpolation_data}, where the in-depth energy deposition profiles $dE/dx(E_k,\theta_k;x)$ are shown for two combinations of $(E_k,\theta_k)$. It is important to highlight that only a fraction of the electron energy $E_k$ is deposited, see the inserts in figures (a)-(b). For the given B-field strength, B-field inclination angle and material composition, this fraction value is dictated not only by the initial energy $E_k$ but also by the impact angle $\theta_k$, for detailed discussion see Refs.\cite{Ratynskaia1, Ratynskaia2}. The interpolation of the whole \textsc{Geant4}-based dataset is illustrated in \,\ref{fig:interpolation_data} (c) where the depth corresponding to $95\%$ of the deposited energy is plotted as function of $(E_k,\theta_k)$.

\subsection{The workflow}

\begin{figure*}[t]
\centering
\begin{tikzpicture}[
    node distance=0.5cm and 1.2cm,
    % --- Shape styles (structure only, no color) ---
    stepbox/.style={rectangle, draw, thick, text width=3.8cm, minimum height=1.4cm,
                    font=\small\sffamily, align=center},
    procbox/.style={rectangle, draw, thick, text width=3.8cm, minimum height=1.4cm,
                    font=\small\sffamily, align=center},
    databox/.style={rectangle, draw, dashed, text width=3.8cm, minimum height=1.4cm,
                    font=\footnotesize, align=left},
    % --- Color styles (composable, single consistent opacity) ---
    colStage/.style={fill=blue!10},
    colProc/.style={fill=orange!10},
    colFinal/.style={fill=red!10},
    arrow/.style={-{Stealth[length=3mm]}, thick},
]
    % --- Stage 1: Simulation ---
    \node (s1) [stepbox, colStage] {RE simulation + Wall collisions};

    % --- Processing actions ---
    \node (pdata) [procbox, colProc, right=of s1] {
        \textbf{Macroparticle data collection}\\
        For macroparticle $k$, extract
        $w_k, \text{wid}_k, \mathbf{p}_k, t_k, E_k$
    };
    \node (agg) [procbox, colProc, right=of pdata] {
        \textbf{Mesh Aggregation}\\
        Group data by Wall ID\\
        Compute $A, \mathbf{n}, \theta_k$
    };
    % --- Stage 2: Data processing (triangle level) ---
    \node (s2) [stepbox, colStage, below=of agg] {
        \textbf{Data Processing}\\
        \textit{(Triangle Level)}
    };
    \node (interp) [procbox, colProc, left=of s2] {
        Interpolate $\frac{dE}{dx}(E_k, \theta_k, x)$, for all $k$\\
        Construct $S(x,t)$
    };
    % --- Final governing equation ---
    \node (1deq) [stepbox, colFinal, left=of interp] {
        \textbf{1D heat diffusion equation}\\
        Temperature evolution $T(x,t)$
    };
    % --- Connections ---
    \draw [arrow] (s1) -- (pdata);
    \draw [arrow] (pdata) -- (agg);
    \draw [arrow] (agg) -- (s2);
    \draw [arrow] (s2) -- (interp);
    \draw [arrow] (interp) -- (1deq);
\end{tikzpicture}
\caption{The surrogate model workflow; the kinetic data from RE macro-particles is aggregated and mapped onto the wall mesh elements to provide volumetric energy source to the 1D heat diffusion solver for wall temperature evolution.}
\label{fig:workflow}
\end{figure*}
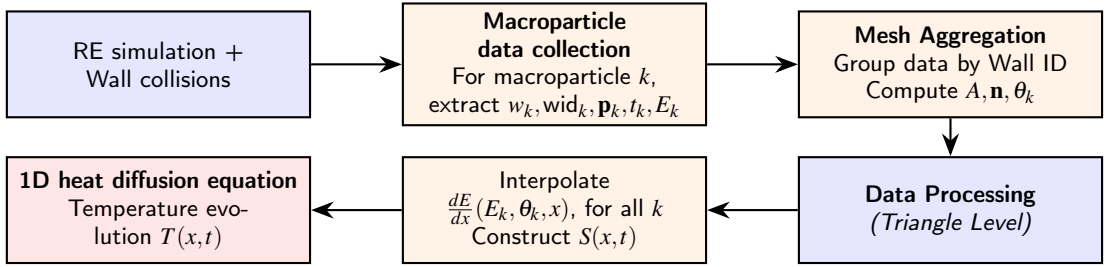

While FIREWALL is intrinsically designed to be used with RE wall load data from arbitrary codes and can be applied to arbitrary geometries, we base our tests and demonstrations on RE wall load data created with the JOREK code~\cite{RE_fluid}. For JOREK, a wall collision model suitable for producing this data has recently been introduced~\cite{Bandaru_2024, Bergstrom2024}. The data used is based on RE fluid simulations with JOREK that were post-processed with a test particle scheme~\cite{Sommariva_2018} which can be used either with full orbit markers (gyro-motion) or guiding center markers (averaging of the gyro-motion). To obtain the most accurate impact angles of REs onto wall elements, we use full orbit particles. Each macro-particle carries a weight that reflects the number of physical REs that it represents.

The general workflow used in the surrogate model is illustrated in Figure~\ref{fig:workflow}. First, an RE simulation is run and collisions of the REs with the PFCs are determined using the wall collision model. For each macro-particle $k$ striking the wall, its weight $w_k$,  wall element number $\text{wid}_k$ (ID of the triangle with which the particle collided), momentum  $\textbf{p}_k$ and time of intersection $t_k$ are collected. From the momentum, the macro-particle energy $E_k$ is computed from the relativistic formula:
\begin{equation} \label{eq:energy}
    E_k = w_k \Big( \sqrt{(||\textbf{p}_k||c)^2 + (m_0c^2)^2} - m_0c^2 \Big),
\end{equation}
where $m_0$ is the electron mass in Dalton. Afterwards, the macro-particle data is grouped by wall ID. For each mesh triangle, its surface area $A$ as well as its unit normal $\textbf{n}$ are computed and for each deposited macro-particle, the incidence angle $\theta_k$ with respect to the surface is computed according to:
\begin{equation} \label{eq:angle}
\theta_k = \Bigg|\arcsin\Bigg (\frac{<\textbf{p}_k, \textbf{n}>}{||\textbf{p}_k||} \Bigg )\Bigg|.
\end{equation}

The next step concerns the extraction of the energy deposition profile $dE/dx(E_k, \theta_k, x)$ for every macro-particle $k$ from the interpolated \textsc{Geant4} dataset and the reconstruction of the volumetric source term $S(x,t)$ as the cumulative sum of the weighted profiles
\begin{equation} \label{eq:source}
S(x,t) =  \frac{1}{A}\sum_{k=1}^N w_k\cdot \chi_k(t) \cdot \frac{dE}{dx}(E_k,\theta_k,x)\,.
\end{equation}

In the above, in order to translate the RE energy density into the RE power density, the function $\chi_k(t)=[H(t - t_k) - H(t-t_k - \Delta_t)]/\Delta_t$ is introduced signifying that in JOREK the energy loading is enabled by macro-particles coming into the wall element at different times $t_k$ and hence it is intrinsically nonuniform. Here $H$ is the Heaviside step function and $\Delta_t$ is the temporal discretization. 

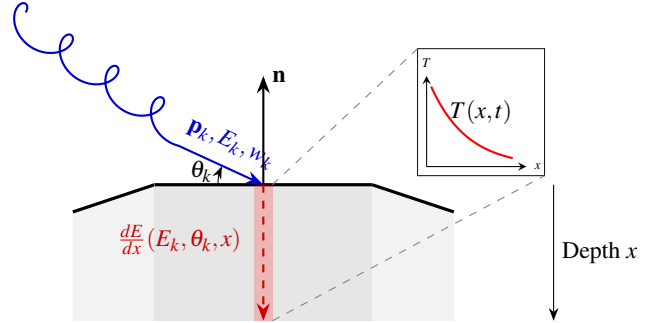
\begin{figure}
\begin{tikzpicture}[
    scale=1.2,
    interface/.style={draw, line width=1.2pt, color=black},
    electron/.style={-Stealth, thick, blue!80!black},
    dimension/.style={thin, Stealth-Stealth, font=\small},
    normal/.style={-Stealth, black, thick},
    plotline/.style={red, thick, smooth}
]
    % --- THE SEGMENTED WALL SURFACE ---
    % Segment 1: Left
    \fill[gray!10] (-2.1, -1.5) -- (-2.1, -0.3) -- (-1.2, 0) -- (-1.2, -1.5) -- cycle;
    \draw[interface] (-2.1, -0.3) -- (-1.2, 0);
    % Segment 2: Middle (Flat Target)
    \fill[gray!20] (-1.2, 0) -- (1.2, 0) -- (1.2, -1.5) -- (-1.2, -1.5) -- cycle;
    \draw[interface] (-1.2, 0) -- (1.2, 0);
    % Segment 3: Right
    \fill[gray!10] (1.2, 0) -- (2.1, -0.3) -- (2.1, -1.5) -- (1.2, -1.5) -- cycle;
    \draw[interface] (1.2, 0) -- (2.1, -0.3);
    % --- RE TRAJECTORY ---
    \coordinate (O) at (0,0);
    \coordinate (surface_ref) at (-1, 0);
    \coordinate (Normal) at (0, 1.2);
    \draw[electron] 
        plot[domain=0:11.75, samples=150, smooth] 
        ({-2.8 + 0.16*\x + 0.18*cos(2*\x r)}, {1.9 - 0.11*\x + 0.18*sin(2*\x r)})
        -- (O) 
        node[midway, sloped, above, font=\small, yshift=2pt] {$\mathbf{p}_k, E_k, w_k$};
    \coordinate (NormalEnd) at (0, 1.2);
    \draw[normal] (O) -- (NormalEnd) node[right, black] {$\mathbf{n}$};
    % --- ANGLE ---
    \coordinate (PreImpact) at (-1.8, 0.8); 
    \pic [draw, Stealth-, "$\theta_k$", angle eccentricity=1.4, angle radius=0.6cm] 
        {angle = PreImpact--O--surface_ref};
    % --- VERTICAL ENERGY DEPOSITION ---
    \begin{scope}
        \clip (-1.2, 0) rectangle (1.2, -1.5);
        \fill[red, opacity=0.2] (-0.1, 0) -- (0.1, 0) -- (0.1, -1.5) -- (-0.1, -1.5) -- cycle;
        \draw[dashed, red!80!black, thick, -Stealth] (0,0) -- (0, -1.5);
    \end{scope}
    \node[red!80!black, anchor=east] at (-0.15, -0.60) {$\frac{dE}{dx}(E_k,\theta_k,x)$};
    % --- 5. INSET TEMPERATURE PLOT ---
    % Drawing a mini-axis
    \begin{scope}[shift={(1.8, 0.2)}]
        % Background for inset
        \fill[white, opacity=0.9, draw=black, thin] (-0.1,-0.1) rectangle (1.3,1.3);
        % Inset Axes
        \draw[->, >=stealth, thin] (0,0) -- (1.1,0) node[right, font=\tiny] {$x$};
        \draw[->, >=stealth, thin] (0,0) -- (0,1) node[above, font=\tiny] {$T$};
        % Temperature Curve (Exponential Decay)
        \draw[plotline] plot[domain=0.05:0.95, samples=50] (\x, {exp(-2.5*\x)});
        \node[font=\small, black] at (0.6, 0.6) {$T(x,t)$};
    \end{scope}
    % Connection from red area to inset
    \draw[gray, thin, dashed] (0.1, 0) -- (1.7, 1.5);
    \draw[gray, thin, dashed] (0.1, -1.5) -- (3.1, 0.1);
    % --- DEPTH AXIS ---
    \draw[dimension, -Stealth] (3.2, 0) -- (3.2, -1.5) node[midway, right] {Depth $x$};
\end{tikzpicture}
\caption{Schematic picture of a wall element with an incident RE.}
\label{fig:incomingRE}
\end{figure}

This source term is then used to calculate the temperature evolution $T(x,t)$ of the mesh element by solving the 1D heat-diffusion equation 
\begin{equation} \label{eq:heat_eq}
    \rho(T)\cdot{c}_{\mathrm{p}}(T)\cdot\frac{\partial T}{\partial t} = \frac{\partial}{\partial x}\left[k(T)\cdot \frac{\partial T}{\partial x}\right] + S(x,t),
\end{equation}
for $(x,t) \in [0,x_d] \times [t_{1}, t_2]$ and with the following initial and boundary conditions:
\begin{equation}
\begin{cases}
 T(x,0) = T_{\text{ini}}(x), \quad x \in [0,x_d] \\
 \displaystyle\frac{\partial T}{\partial x}(0,t) = \displaystyle\frac{\partial T}{\partial x}(x_d,t) = 0, \quad t \in [t_1, t_2].
\end{cases}
\end{equation}
Here, $\rho(T)$, $k(T)$ and $c_{\mathrm{p}}(T)$ are the temperature dependent mass density, thermal conductivity and specific isobaric heat capacity, respectively. For tungsten, we follow the recommendations of Ref.\cite{W_prop}, as also implemented in MEMENTO \cite{Paschalidis2024}. A schematic illustration of an incident RE onto a mesh triangle, with the relevant quantities, is shown in Figure~\ref{fig:incomingRE}.

Equation \ref{eq:heat_eq} is solved using a finite-volume discretization on a bi-uniform depth grid, with a finer resolution near the plasma-facing surface in order to resolve the steep gradients of the volumetric RE energy deposition profiles. The time integration is carried out using an implicit Euler scheme, with the temperature-dependent thermophysical properties resolved via Picard iteration, in which properties are updated from the latest temperature estimate at each iteration until the temperature field converges. Thermal conductivity at cell interfaces is obtained by harmonic averaging in order to ensure consistent heat fluxes across the non-uniform grid. The resulting tridiagonal system is solved with the Thomas algorithm \cite{tridiag} at each iteration. The spatial resolution is chosen not to exceed that of the underlying \textsc{Geant4} energy-deposition profiles.

Finally, this workflow is parallelized with OpenMP by a load distribution based on the mesh triangles. 

\begin{figure*}[t]
    \centering
    \begin{overpic}[width=\textwidth]{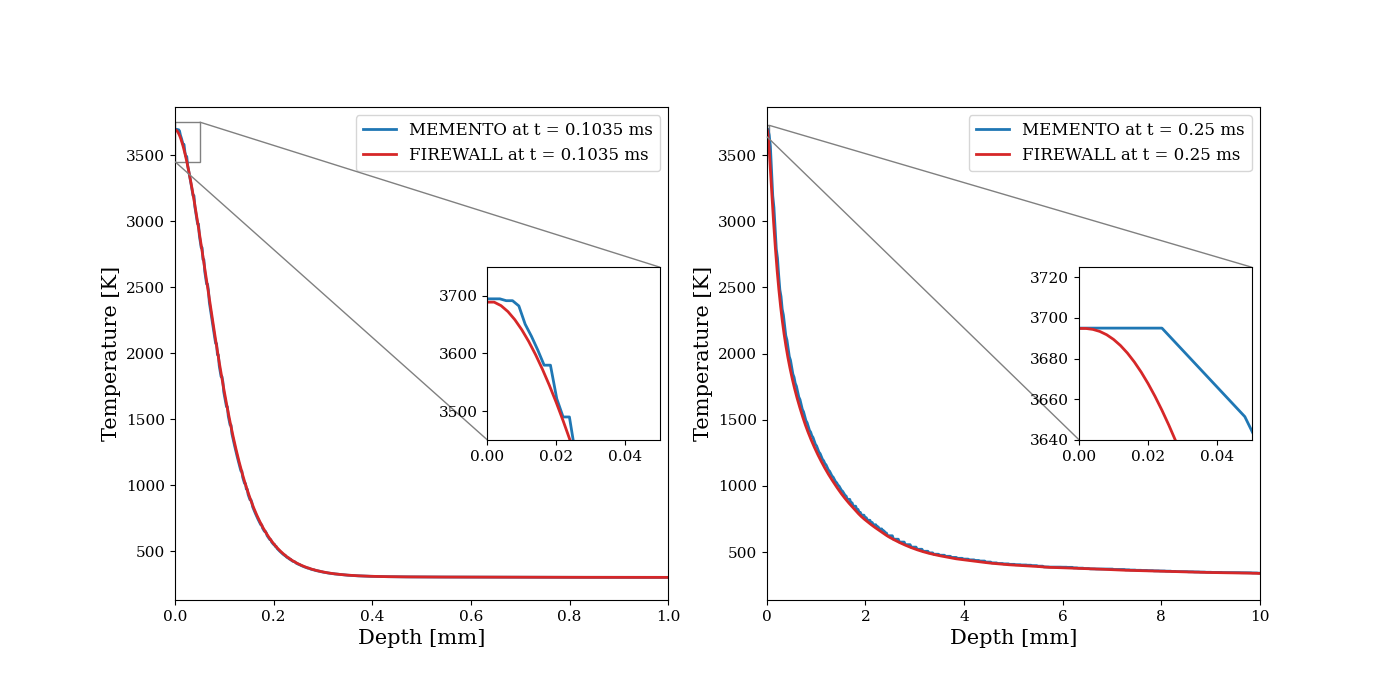}
        \put(20,35){\textbf{a)}}  % Top left
        \put(84,35){\textbf{b)}} % Top right
    \end{overpic}
    \caption{Temperature profiles obtained with the FIREWALL heat solver (red) and MEMENTO (blue), shown at the timestep when MEMENTO reaches melting. Results for the volumetric source in Fig.~\ref{fig:interpolation_data}(a) at the left panel and for the volumetric source in Fig.~\ref{fig:interpolation_data}(b) at the right panel.}
    \label{fig:simulation_results}
\end{figure*}

\section{Benchmarking of the solver against the Geant4-MEMENTO workflow}\label{sec3}

To assess the accuracy of the model and estimate the uncertainties of the FIREWALL calculations, detailed comparison with the high-fidelity \textsc{Geant4}-MEMENTO thermal response workflow~\cite{Ratynskaia1} is performed. First, in Sec.~\ref{subsec:bench_Tsolver}, aiming to isolate effects of different assumptions, the 1D heat solver benchmarking is performed on a simplistic geometry with given volumetric inputs. Next, the full workflow is benchmarked; the accuracy of the reconstruction of the energy density profiles is assessed (Sec.~\ref{subsec:bench_Energy}) and comparison of the predictions for the final temperature profile and the onset of melting is carried out (Sec.~\ref{subsec:bench_Temp}).

It is reminded that MEMENTO is a macroscopic melt motion code that solves the heat and phase transfer problem coupled with the incompressible Navier–Stokes equations within the shallow water approximation together with the current propagation equations on a domain that features a time-evolving deforming solid-plasma interface\,\cite{Ratynskaia3,Ratynskaia4}. Here, only the thermal module of MEMENTO is considered which solves the three dimensional heat and phase transfer problem for an evolving domain owing to vaporization\,\cite{Ratynskaia1} by utilizing non-uniform adaptive meshing along with sub-cycling in time\,\cite{Paschalidis2024}.

\subsection{Benchmarking of the heat solver}\label{subsec:bench_Tsolver}
We compare the results obtained with the 1D heat solver implemented in FIREWALL against those obtained with MEMENTO. The cases studied correspond to those illustrated in Fig.\,\ref{fig:interpolation_data}, featuring shallow energy deposition for the 1\,MeV, 2$^{\circ}$ pair and deep energy deposition for the 25\,MeV, 10$^{\circ}$ pair. 

The simulations are carried out in simplified geometry; a $1\times1\times24\,\mathrm{mm}^3$ tungsten slab of $T_{\text{ini}} = 300$\,K uniform initial temperature. The loading is uniform in space and time over a $t_{\text{load}}=1$\,ms duration. The FIREWALL runs are performed with the following parameters: $\Delta_t = 10^{-7}$\,s, $t_1 = 0$, $t_2 = t_{\text{load}}=10^{-3}$\,s, $\delta_{x}=2\times 10 ^{-6}$\,m, $A=10^{-6}\,\text{m}^2$, $x_d = 24 \times 10^{-3}$\,m. Here, $\delta_x$ is the grid size which should be smaller than or equal to the in-depth resolution of the \textsc{Geant4} energy deposition profiles. The total energy loaded, $E_{\mathrm{load}}$, is 10\,J in case (a) and 40\,J in case (b). These values are chosen in order to ensure a prompt melting for each of the profiles and negligible surface cooling.

In the FIREWALL runs, the continuous loading is approximated by loading one macroparticle for a duration of $t_{\text{load}}$. In such case the expression (\ref{eq:source}) simplifies to 
\begin{equation}
S(x) = \frac{1}{A}\frac{E_{\text{load}}}{ \int^{x _d}_0 \frac{dE}{dx}(E_k, \theta_k, x) dx} \frac{1}{t_{\text{load}}} \frac{dE}{dx}(E_k,\theta_k,x), \
\label{source_uniform}
\end{equation}
where the macroparticle weight $w$ is expressed by the second fraction in the equation, accounting for energy losses.

The comparison results are presented in Fig.\ref{fig:simulation_results} at the time when melting is reached in the MEMENTO simulations; at $t_{\text{a}} = 0.10$ ms in case (a) and at $t_{\text{b}} = 0.25$ ms in case (b). The relative $L_2$ error of the temperature profiles is around $0.4\%$ for scenario (a) and  about $2.9 \%$ in (b). The accuracy in predicting the onset of melting is also satisfactory,  with FIREWALL reaching melting some microseconds later, yielding a relative error in time of $\sim0.27\%$ in (a) $\sim0.067\%$ in (b).

\begin{figure*}[!t]
    \centering
    \begin{overpic}[width=0.99\linewidth]{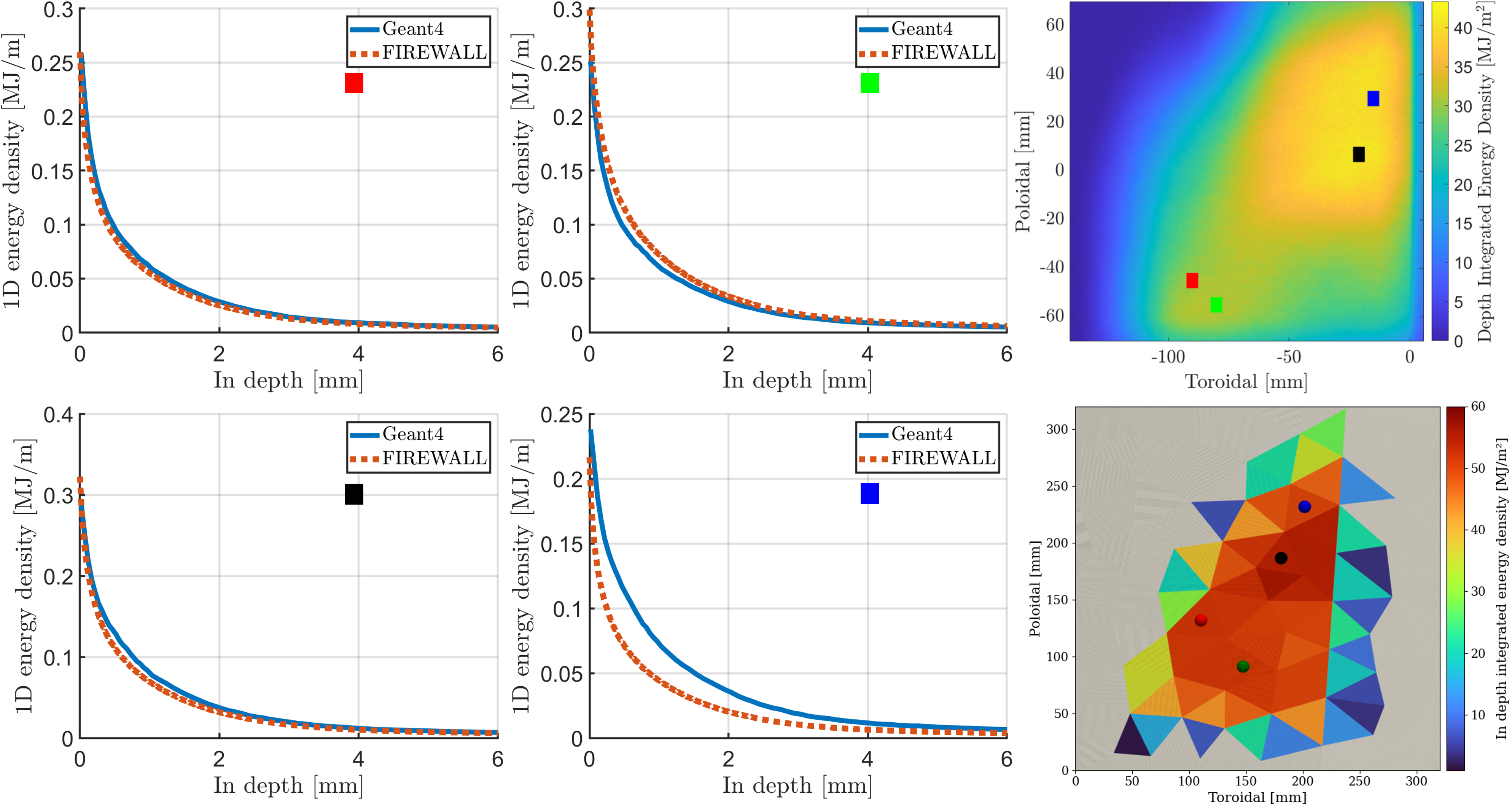}
        \put(10,48){\textbf{\hbox{\kern4pt\textcolor{black}{\textbf{a)} }}}}
        \put(10,18){\textbf{\hbox{\kern4pt\textcolor{black}{\textbf{b)} }}}}
        \put(45,48){\textbf{\hbox{\kern4pt\textcolor{black}{\textbf{c)} }}}}
        \put(45,18){\textbf{\hbox{\kern4pt\textcolor{black}{\textbf{d)} }}}}
        \put(72,48){\textbf{\hbox{\kern4pt\textcolor{white}{\textbf{e)} }}}}
        \put(72,18){\textbf{\hbox{\kern4pt\textcolor{black}{\textbf{f)} }}}}
    \end{overpic}
    \caption{(a-d) Comparison of the energy density reconstruction in FIREWALL against the results of 3D \textsc{Geant4} simulations, for scenario J1. The in-depth energy density profiles are compared for four different locations in the panel, as assigned by the color coding. (e) Depth-integrated energy density map from the 3D \textsc{Geant4} simulations. (f) Depth-integrated energy density map from the FIREWALL simulations.}
    \label{fig:Energy_J1}
\end{figure*}

\begin{figure*}[!t]
    \centering
        \begin{overpic}[width=0.93\linewidth]{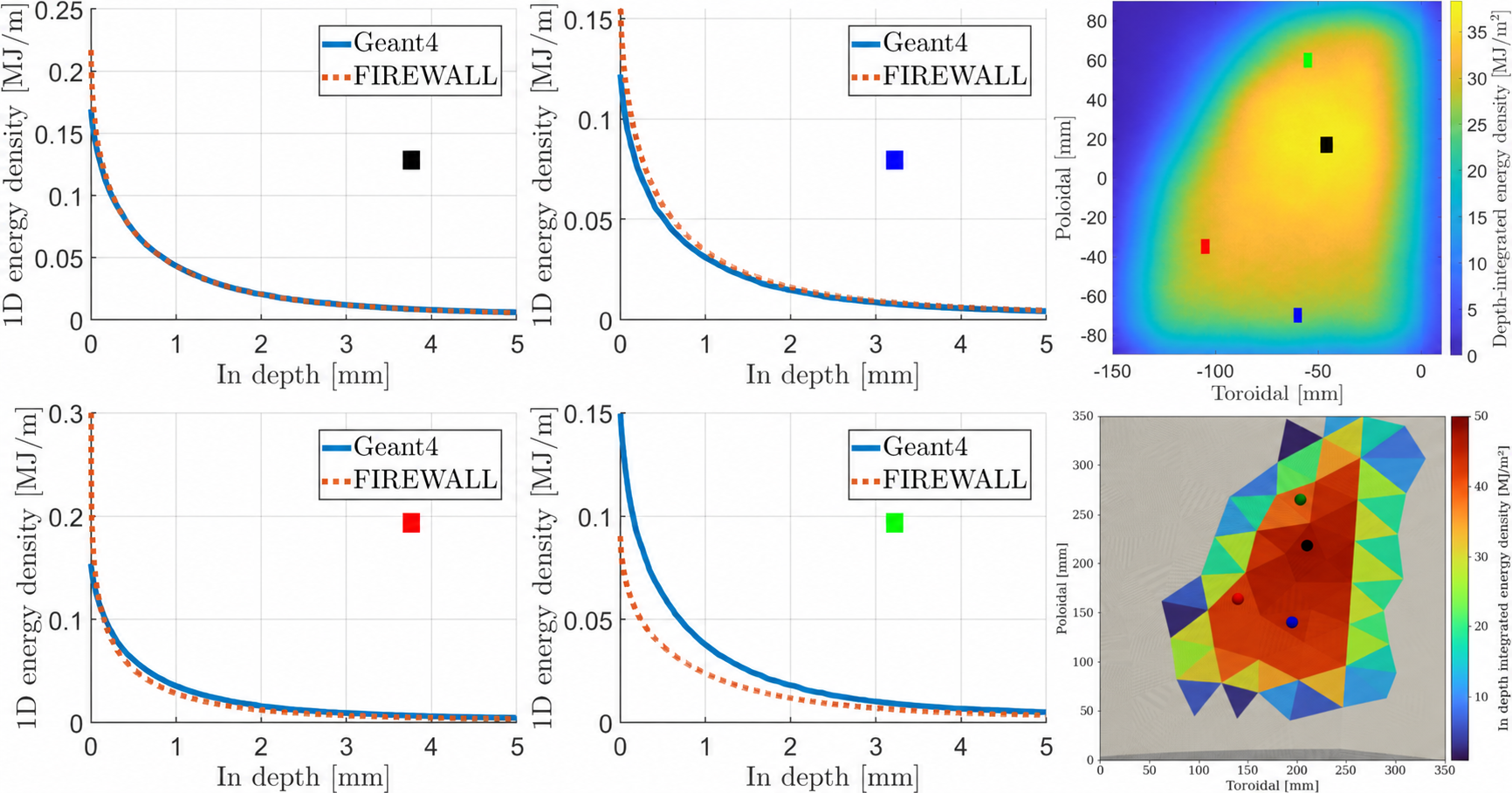}
        \put(8,50){\textbf{\hbox{\kern4pt\textcolor{black}{\textbf{a)} }}}}
        \put(8,20){\textbf{\hbox{\kern4pt\textcolor{black}{\textbf{b)} }}}}
        \put(45,50){\textbf{\hbox{\kern4pt\textcolor{black}{\textbf{c)} }}}}
        \put(45,20){\textbf{\hbox{\kern4pt\textcolor{black}{\textbf{d)} }}}}
        \put(74,50){\textbf{\hbox{\kern4pt\textcolor{white}{\textbf{e)} }}}}
        \put(74,20){\textbf{\hbox{\kern4pt\textcolor{black}{\textbf{f)} }}}}
    \end{overpic}
    \caption{(a-d) Comparison of the energy density reconstruction in FIREWALL against the results of 3D \textsc{Geant4} simulations, for scenario J4. The in-depth energy density profiles are compared for four different locations in the panel, as assigned by the color coding. (e) Depth-integrated energy density map from the 3D \textsc{Geant4} simulations. (f) Depth-integrated energy density map from the FIREWALL simulations.}
    \label{fig:Energy_J4}
\end{figure*}

\subsection{Accuracy of energy density reconstruction} \label{subsec:bench_Energy}

The aim of this section is to test FIREWALL's accuracy in reconstructing in-depth energy density profiles from the discrete database, i.e., the focus lies exclusively on the uncertainties in the source term $S$ given by Eq.~(\ref{eq:source}).  

Two RE beam termination scenarios on the ITER first wall, considering a high current of $14.6\,\rm MA$, are chosen. These scenarios had been simulated with the JOREK RE fluid model\,\cite{RE_fluid}, as detailed in Refs\,\cite{Bandaru_2024, Bergstrom2024} and post-processed with the particle tracker model to retrieve the wall loading. This JOREK output had been employed in high-fidelity thermal response simulations\,\cite{Ratynskaia1}, where the studied scenarios were labeled as J1 and J4, a notation also followed hereafter. 

The comparison is performed for the most loaded panel in each of the two JOREK scenarios. Namely, in J1, JOREK assumes a simplistic mono-energetic RE distribution at 26\,MeV and an initial pitch angle such that $p_\parallel/p_{\rm total} = 0.99$, where $p_\parallel$ is the particle momentum along the magnetic field line and $p_{\rm total}$ is the total particle momentum. In J4, a more realistic avalanche distribution (with a mean of 18\,MeV) given analytically in Ref.\cite{Embreus_2018} is employed, coupled to a broader pitch angle distribution. These two scenarios offer the possibility to check separately how accurate the reconstruction of the energy deposition profile is in FIREWALL, separating the effect of the $\theta_k$ distribution (see the constant $E_k$ in J1) from a more convoluted scenario with combined $E_k,\theta_k$ distributions (in J4), where the RE initial energy ranges between about 1 and 150\,MeV, while the impact angle lies between 0 and 60$^{\circ}$. For an ilustration of the impact angle distributions in J1,\,J4 refer to Fig.2 of Ref.\cite{Ratynskaia1}.

The results of the comparison are shown in Figs.\ref{fig:Energy_J1}-\ref{fig:Energy_J4}(a-d), where the in-depth energy deposition profiles reconstructed by FIREWALL are plotted along with the results of the 3D \textsc{Geant4} simulations.  For each panel, the results correspond to four wall elements (triangles), including the most loaded one (marked in black). In order to facilitate a comparison, tangential averaging of the \textsc{Geant4} maps is performed, over $20\times20$ cells in the poloidal and toroidal directions. In fact, due to a very fine tangential mesh ($\sim100\,\mu$m) employed in the \textsc{Geant4} simulations, single-point in-depth profiles are noisy, while a tangential average enables a far more meaningful comparison, without losing information on the in-depth gradients. For a quick overview of the panel loading, also the depth integrated energy density (in units of energy per area) is plotted from \textsc{Geant4} (e) and FIREWALL (f) simulations for the J1 (Fig.\ref{fig:Energy_J1}) and J4 (Fig.\ref{fig:Energy_J4}) scenarios. Finally, we note that the Monte Carlo simulations are time independent while loading in JOREK is nonuniform in time, see Eq.(\ref{eq:source}), hence in the FIREWALL case all plots in Figs.\ref{fig:Energy_J1} - \ref{fig:Energy_J4} correspond to the time at which the last macroparticle impacted the panel.

The performance of FIREWALL in reproducing the energy deposition can be assessed via: \textbf{(i)} its ability to resolve the spatial structure of the loading and \textbf{(ii)} its accuracy in reconstructing in-depth energy deposition profiles. These will be affected by three-dimensional effects, accounted for in the \textsc{Geant4} runs but not in FIREWALL, as well as by the completeness of the underlying \textsc{Geant4} database and the interpolation schemes implemented in FIREWALL.

In the J1 scenario, the overall level of agreement is satisfactory. The wall element receiving the highest energy load is located near the most affected $16\times16~\text{mm}^2$ region identified in the \textsc{Geant4} simulations. For three out of the four analyzed triangular elements, colored by red, black, and green in panels (a-c) of Fig.~\ref{fig:Energy_J1}, the reconstructed in-depth profiles closely match the \textsc{Geant4} results, with discrepancies in total deposited energy below 10\%. A more pronounced deviation is observed for the element marked in blue (d): here, FIREWALL predicts a steeper profile compared to the \textsc{Geant4} result, with a lower peak energy density and a total energy discrepancy of the order of 20\%. 

In the J4 scenario, where the RE beam is characterized by an energy distribution in addition to the impact angle distribution, the level of agreement has slightly deteriorated. Although FIREWALL captures the main features of the spatial energy distribution, such as the presence of a centrally located highly loaded region, appreciable discrepancies arise in the in-depth profiles. In particular, for the element in green the overall discrepancy between profiles is of the order of 20\%. For the other elements, the profiles match closely in-depth, but near the surface FIREWALL predicts peak energy densities that are up to a factor of 1.5 higher than those from \textsc{Geant4}.

\begin{figure*}[!t]
    \centering
    \begin{overpic}[width=0.4\linewidth]
        {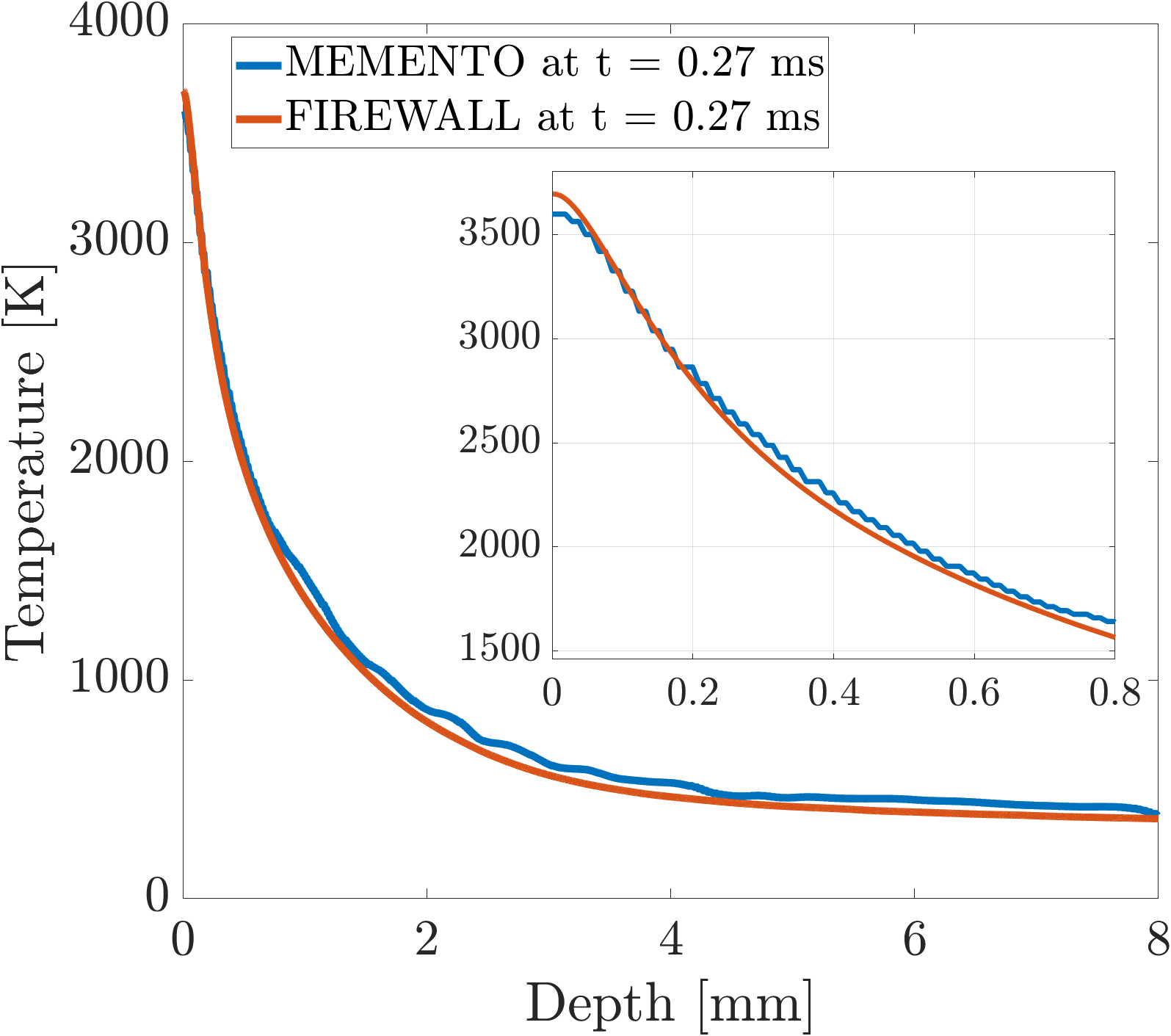}
        \put(86,82){\textbf{a)}}
    \end{overpic}
    \hspace{0.01cm}
    \begin{overpic}[width=0.4\linewidth]
        {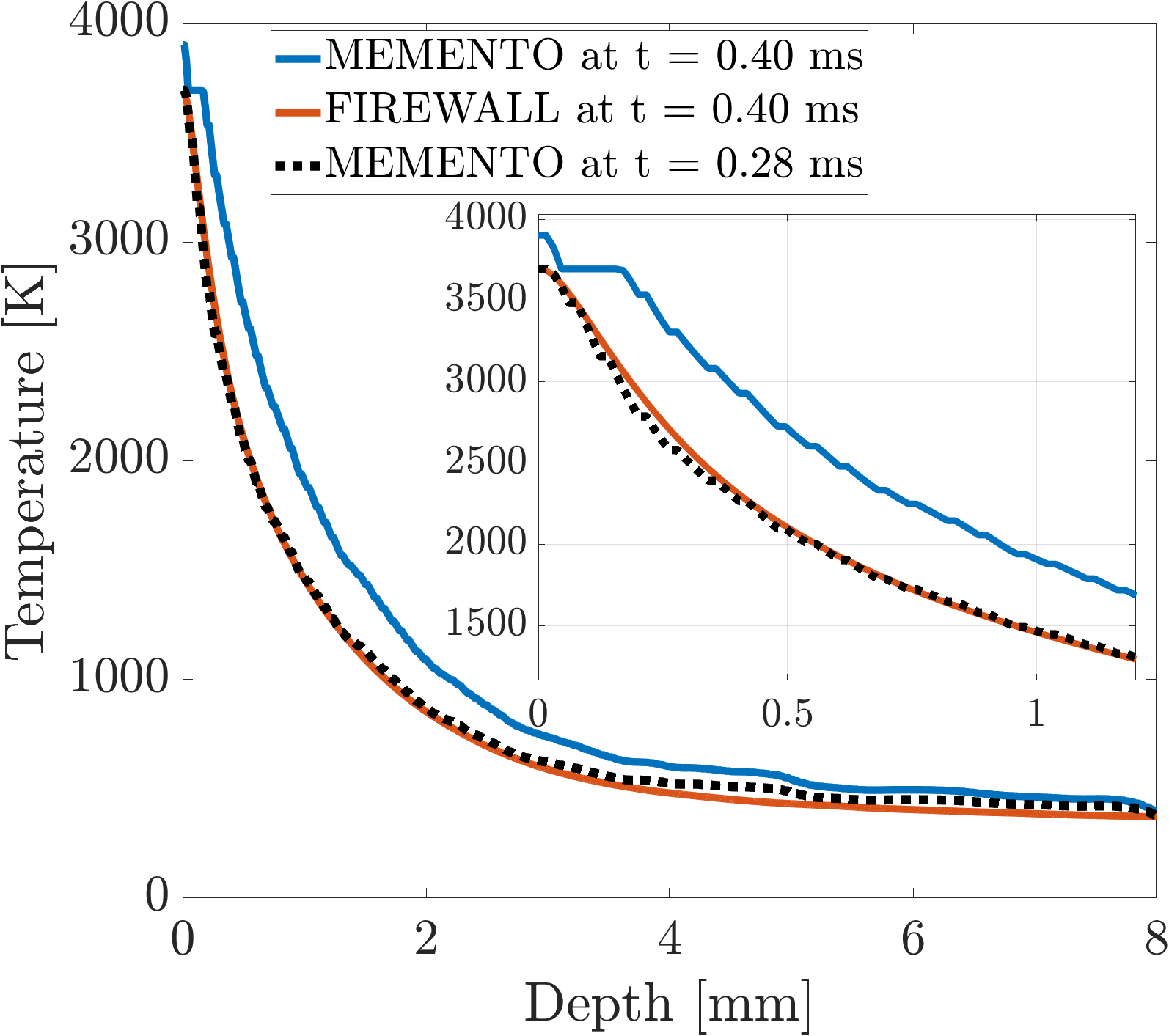}
        \put(86,82){\textbf{b)}}
    \end{overpic}
    \medskip
    \begin{overpic}[width=0.4\linewidth]
        {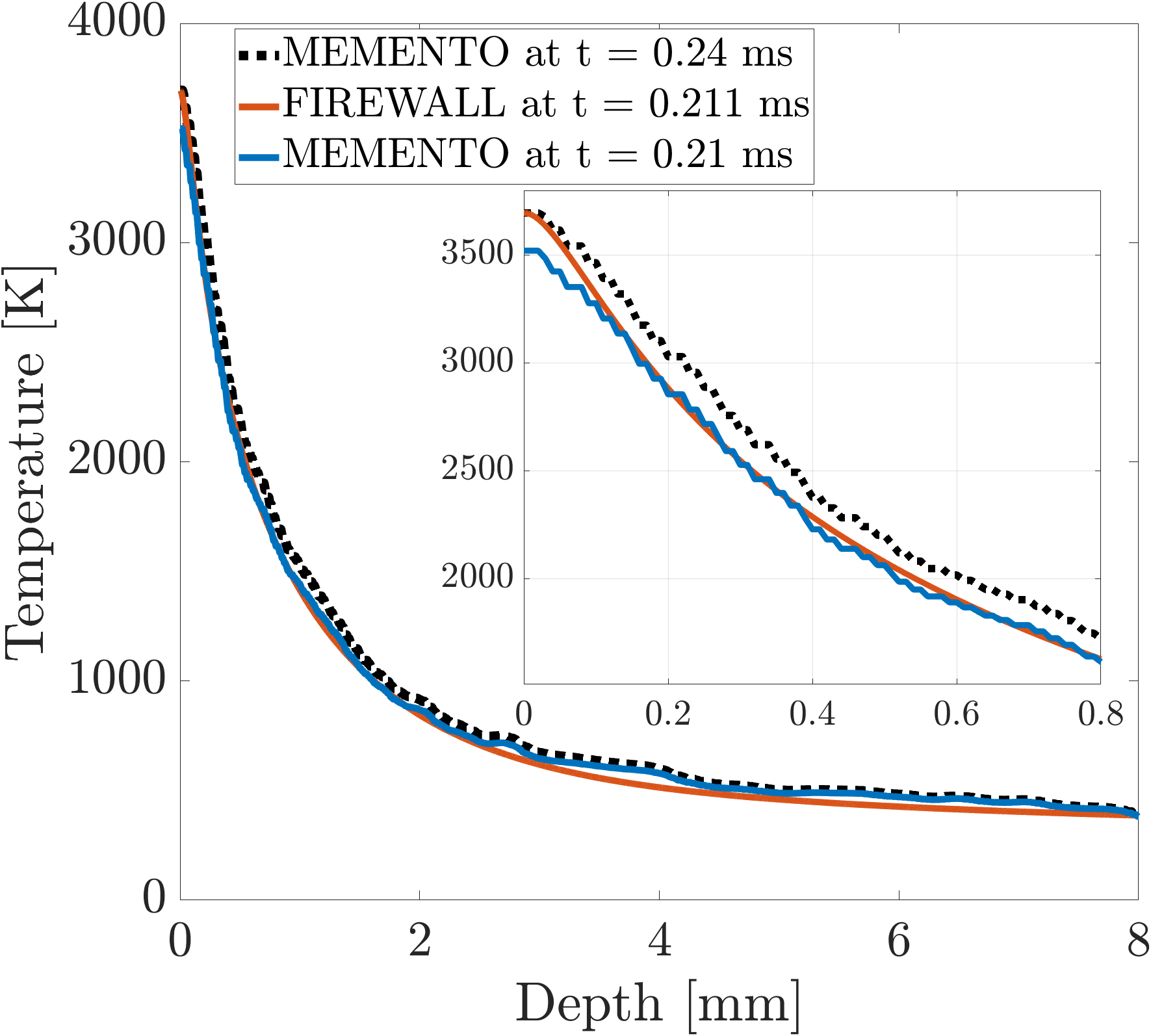}
        \put(86,82){\textbf{c)}}
    \end{overpic}
    \hspace{0.01cm}
    \begin{overpic}[width=0.41\linewidth]
        {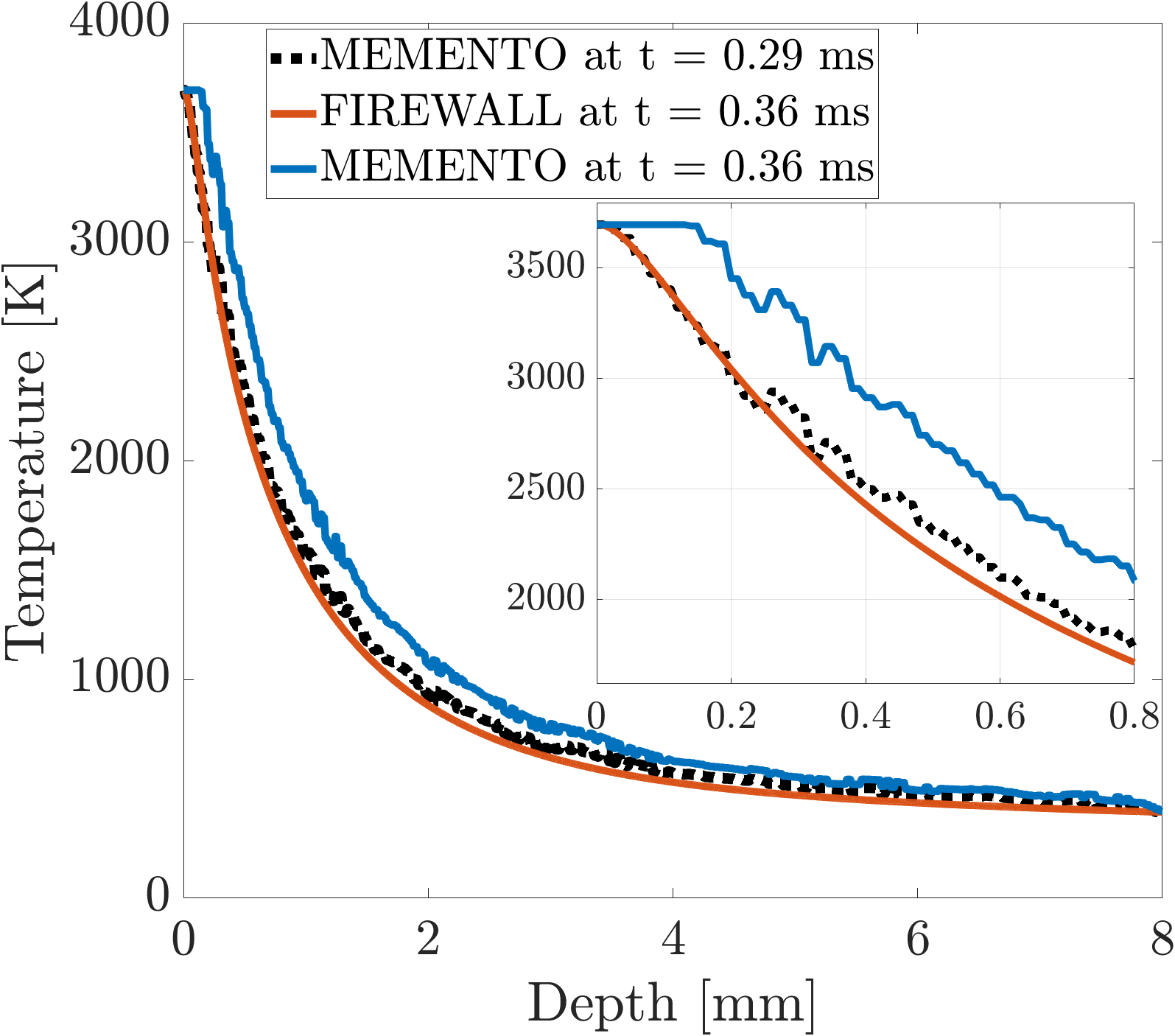}
        \put(86,82){\textbf{d)}}
    \end{overpic}
    \caption{Comparison of the in-depth temperature profiles between FIREWALL and MEMENTO for J1 (a--b) and J4 (c--d).}
    \label{fig:Temperature_comparison}
\end{figure*}

\subsection{Accuracy of the in-depth PFC temperature profile}\label{subsec:bench_Temp}

Now that the accuracy of the in-depth energy deposition reconstruction has been assessed, we turn to a comparison of the resulting thermal response. It is worth pointing out, that there is an effect of the temporal dependence of the source. In particular, in JOREK, the loading at every time depends on the macroparticles impacting a particular wall element. Here, to facilitate the comparison, similar to the MEMENTO simulations of Ref.\cite{Ratynskaia1}, the loading in FIREWALL is also performed under the assumption of uniformity over the loading duration (Eq.(\ref{source_uniform})).

The ITER first wall features castellated tungsten armor that comprises 'small' tiles \cite{PITTS}. Here, following worse-case scenario thermal calculations of Ref.\cite{Ratynskaia1}, we focus on a single tile with dimensions of $16\times16$\,mm$^2$. Such a tile is identified from the corresponding \textsc{Geant4} map by extracting the $16\times16$ mm$^2$ area centered around the specific region of interest (marked by color in Figs~\ref{fig:Energy_J1} (e,f) and ~\ref{fig:Energy_J4} (e,f)). The resulting energy map is then converted into a power density map by assuming a uniform loading in time, and subsequently used in MEMENTO as the source term of the heat equation.

In MEMENTO, the Enhanced Heat Flux (EHF) panels are modeled with a first W layer followed by a 7\,mm Cu-Cr-Zr layer and a simplified hyper-vapotron shape, where the teeth profile is aligned to the tile’s toroidal center\,\cite{Ratynskaia1}, see also Fig.1 therein. The implemented boundary conditions include: thermally insulated sides facing the gaps, vaporization cooling and thermal radiation cooling from the free surface as well as convection cooling flux at the CuCrZr - coolant interface.  

FIREWALL is designed to rapidly assess the potential for RE-induced damage over large vessel areas, rather than to predict the highly detailed thermal response of the first wall. To this end, the presently implemented simplified model does not account for phase transitions and the growth of molten volume during loading. The model features Neumann boundary conditions and considers uniform material. However, given the tool's validity range up to the melting point and also given the short loading times of interest, the contributions of vaporization, thermal radiation and active cooling remain marginal.

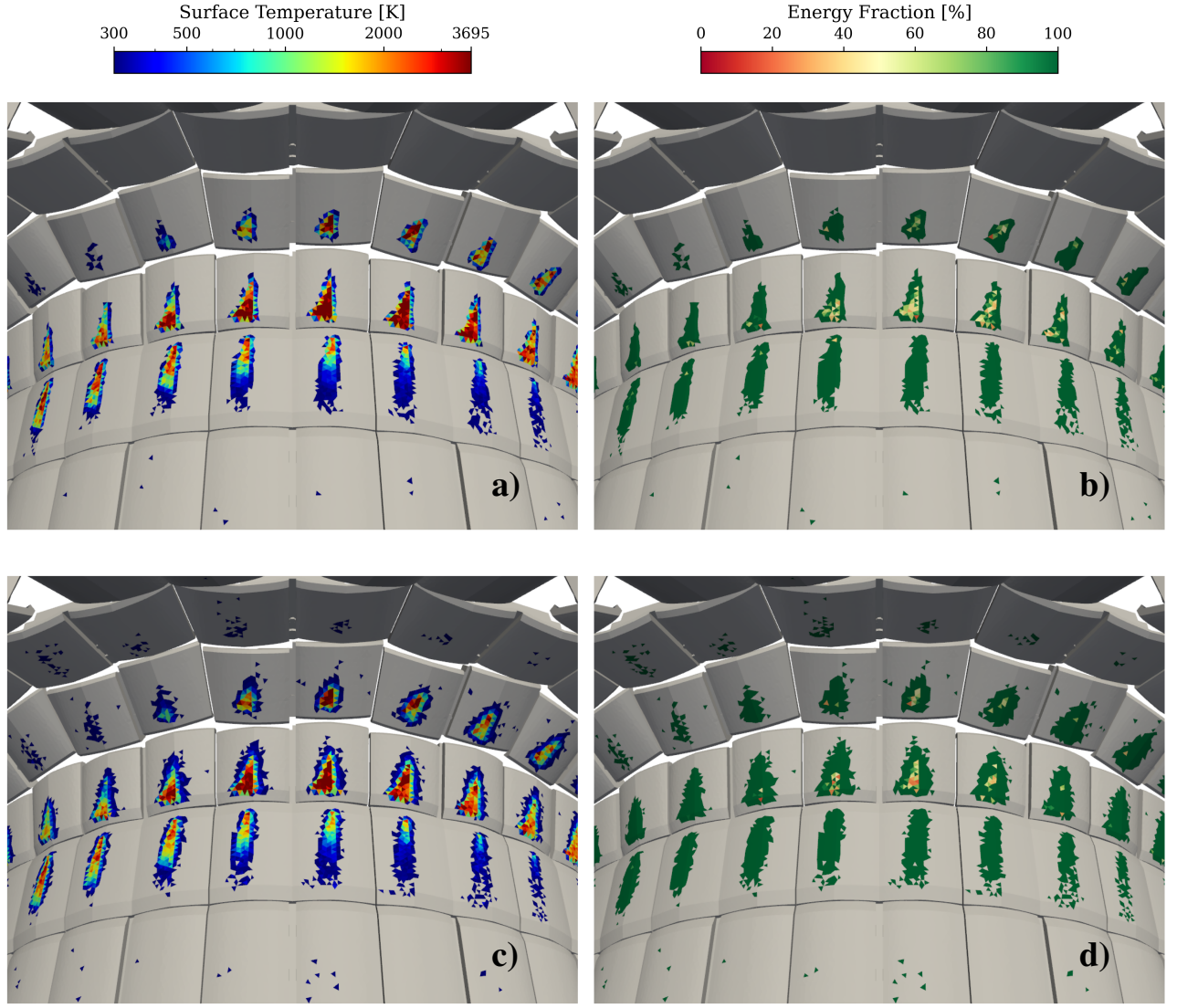
\begin{figure*}[t]
    \centering
    % width=\textwidth ensures it fills the horizontal space of both columns
    \begin{overpic}[width=0.95\textwidth]{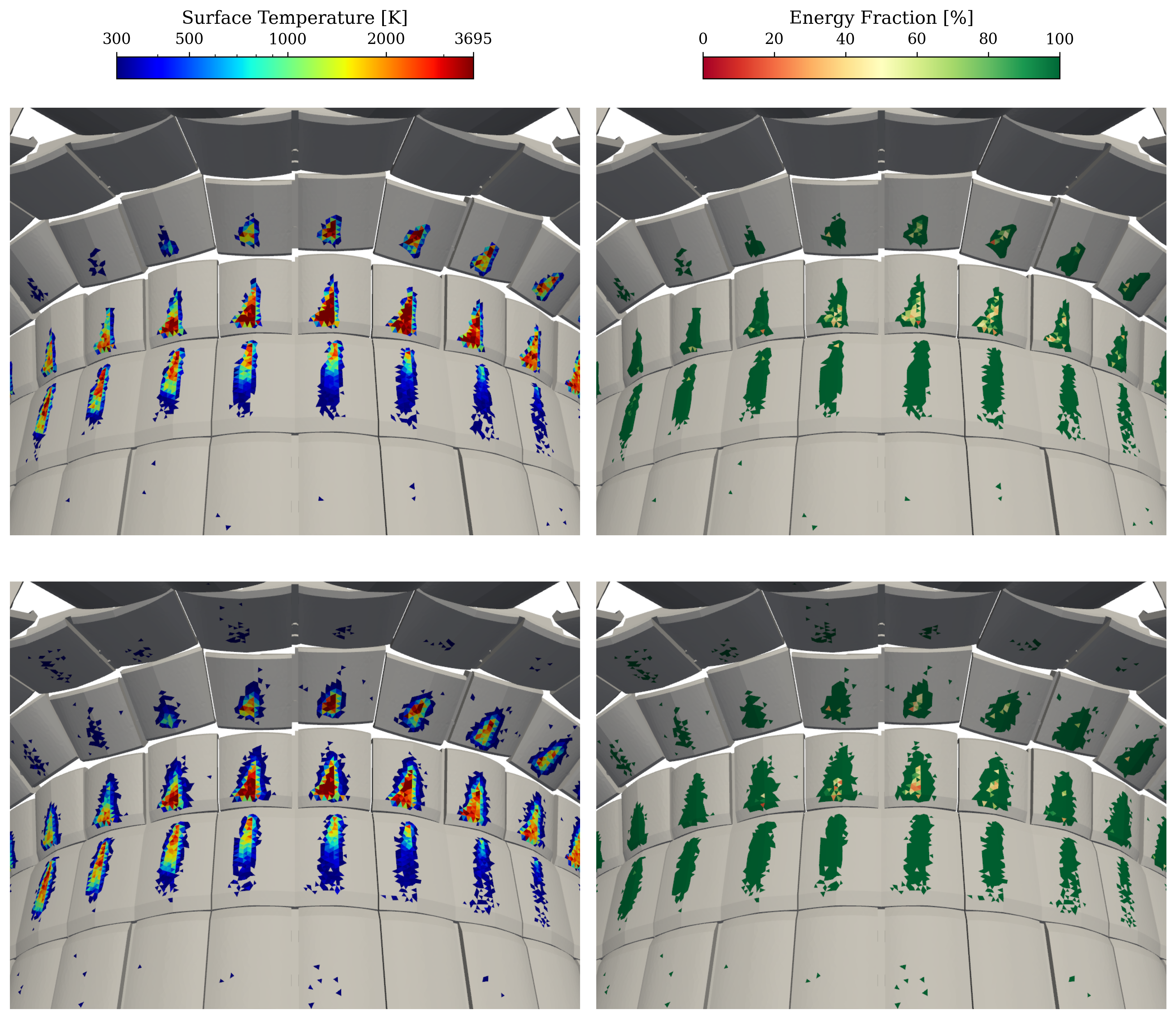}
        \put(42,44){\Large\textbf{a)}}   % top-left subplot: J2 temperature
        \put(92,44){\Large\textbf{b)}}  % top-right subplot: J2 energy fraction
        \put(42,4){\Large\textbf{c)}}   % bottom-left subplot: J4 temperature
        \put(92,4){\Large\textbf{d)}}  % bottom-right subplot: J4 energy fraction
    \end{overpic}
    \caption{FIREWALL results for the thermal response of the full ITER vessel to RE incidence with input from the JOREK simulations for scenarios J1 (top) and J4 (bottom). 3D view of the central column with the temperatures cropped at the melting point of tungsten (left) and with the energy fraction as defined in the main text (right). The plots are taken with the toroidal angle set to 130 degrees for both cases.}
    \label{fig:3d_thermo}
\end{figure*}

\begin{figure*}
    \centering
    % width=\textwidth ensures it fills the horizontal space of both columns
    \includegraphics[width=\textwidth]{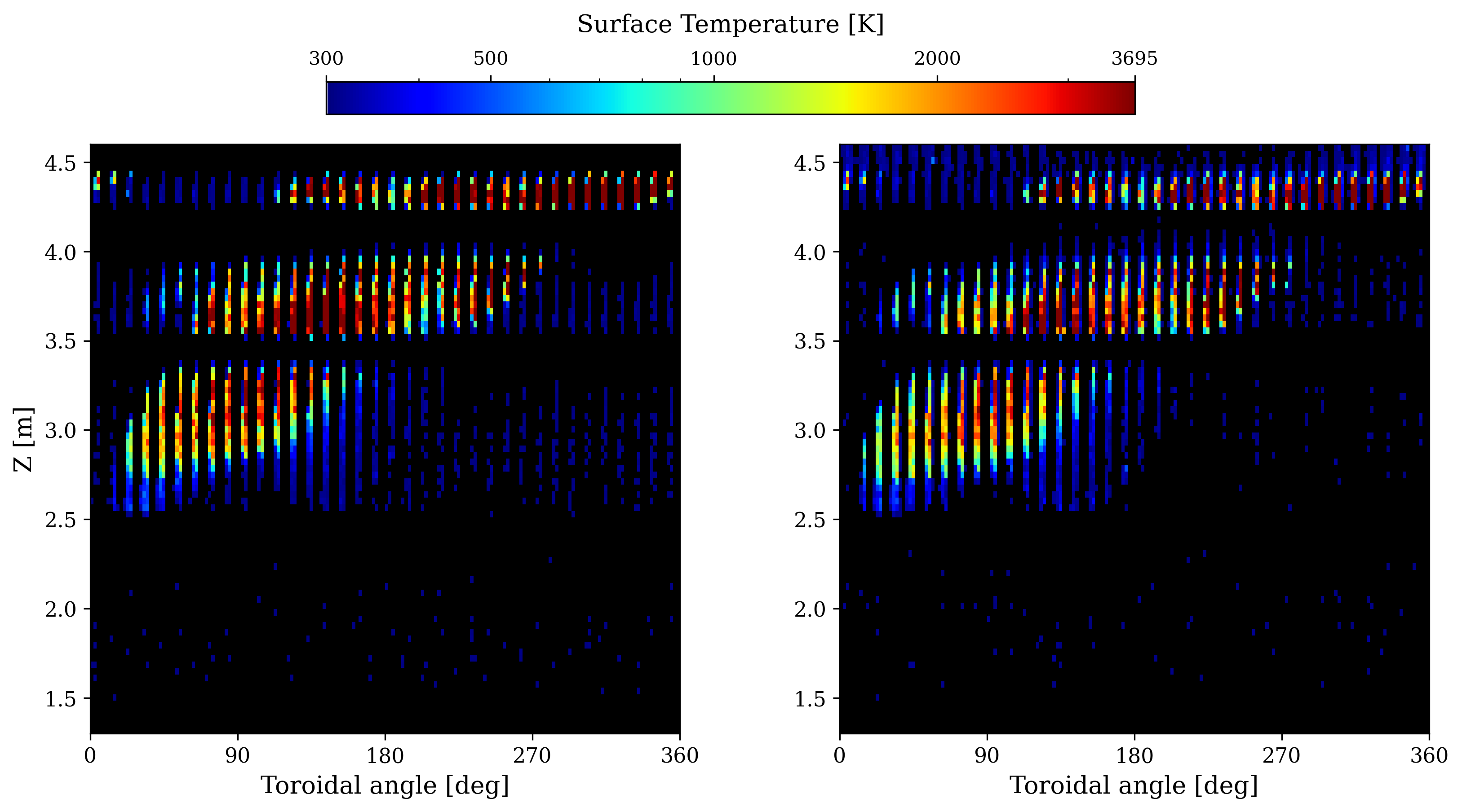}
    \caption{Toroidal projection of the temperatures for JOREK simulations J1 and J4.}
    \label{fig:toroidal_proj}
\end{figure*}

Given these setups, we compare the temperature profiles predicted by FIREWALL for the triangular wall elements corresponding to the EHF tiles simulated with MEMENTO. Fig.~\ref{fig:Temperature_comparison} presents the results for the scenarios J1 (Fig.~\ref{fig:Temperature_comparison}(a,b)) and J4 (Fig.~\ref{fig:Temperature_comparison}(c,d)). The FIREWALL profiles are shown at the first time step at which the melting temperature is reached. The MEMENTO profiles are evaluated at the tile centers and shown at the same times and, where relevant, at the onset of melting in MEMENTO. For both J1 and J4, two locations on the first-wall panel are considered, representing different levels of agreement in the reconstructed energy-deposition profiles. Fig.~\ref{fig:Temperature_comparison}(a,c) correspond to the most heavily loaded wall element, marked in black in Fig.~\ref{fig:Energy_J1} and Fig.~\ref{fig:Energy_J4} , for which the energy-deposition profiles agree closely (Fig.~\ref{fig:Energy_J1}(b), Fig.~\ref{fig:Energy_J4}(a)). Fig.~\ref{fig:Temperature_comparison}(b,d) corresponds to a more peripheral element marked in blue in Fig.~\ref{fig:Energy_J1}  and green in Fig.~\ref{fig:Energy_J4} , for which the agreement is poorer (Fig.~\ref{fig:Energy_J1}(d), Fig.~\ref{fig:Energy_J4}(d)).

For the most heavily loaded element in the J1 scenario, the temperature profiles agree closely: both models predict melting onset at approximately 0.27\,ms, corresponding to about 30\% of the loading duration, and the relative difference between the profiles remains below 5\%. For the peripheral element, MEMENTO predicts an earlier onset of melting than FIREWALL, at 0.28 ms rather than 0.40 ms. This difference is consistent with the discrepancies in the deposited-energy profiles. The MEMENTO profile at 0.40 ms, shown by the solid blue curve, indicates that melting has already progressed a few hundred micrometers below the surface. At this time, the relative difference between the temperature profiles reaches approximately 25\%.

For the most heavily loaded element in the J4 scenario, illustrated in Fig.\ref{fig:Temperature_comparison}(c), FIREWALL predicts an earlier onset of melting than MEMENTO, at 0.21\,ms rather than 0.24\,ms. This behavior is consistent with the higher near-surface peak in the energy density reconstructed by FIREWALL (Fig.\ref{fig:Energy_J4}(a)). At 0.21\,ms, the MEMENTO profile, shown by the solid blue curve, remains below the melting temperature, with a maximum of approximately 3500\,K, about 5\% lower than the maximum predicted by FIREWALL. For the peripheral element, shown in Fig.\ref{fig:Temperature_comparison}(d), MEMENTO predicts an earlier onset of melting than FIREWALL, at 0.29\,ms rather than at 0.36\,ms. This difference is consistent with the discrepancies in the deposited energy profiles (Fig.\ref{fig:Energy_J4}(d)). At 0.36~ms, the MEMENTO profile, shown by the solid blue curve, exhibits a near-surface plateau at the melting temperature, indicating that melting has already progressed below the surface, while the sub-surface temperatures are appreciably higher than those predicted by FIREWALL.

\section{Application to a full ITER domain simulation}
\label{sec4}

To demonstrate the unique capabilities of FIREWALL in a production-scale application, we apply it to the full ITER-domain simulations J1 and J4 described in Refs.\cite{Bergstrom2024, Bandaru_2024}. The initial temperature of all wall elements is set to 300 K. 
Fig.~\ref{fig:3d_thermo} shows the resulting surface-temperature distributions together with the corresponding energy fractions for the two scenarios. Here, the energy fraction is defined as the cumulative energy deposited up to the time at which a wall element first reaches the tungsten melting temperature, normalized to the total energy deposited in that element by the end of its loading. 

This representation provides information beyond a simple map of the surface wall temperature. For elements that remain below the melting point, the energy fraction corresponds to 100$\%$ and the reported temperature is that reached at the end of loading. Such elements in Fig.~\ref{fig:3d_thermo} (a) and (c) correspond to the dark-green areas plots (b) and (d), respectively. In contrast, under more intense loading, the melting point is reached after only a part of the total energy has been deposited, as indicated by the red and yellow areas in Fig.~\ref{fig:3d_thermo} (b) and 8(d). These wall segments will continue to receive substantial energy after the onset of melting, beyond the validity range of the present FIREWALL thermal model, and therefore require subsequent analysis with high-fidelity tools accounting for phase change and melt evolution.

Finally, Fig.~\ref{fig:toroidal_proj} presents toroidal projections of the surface-temperature distributions, providing a compact overview of the spatial extent and localization of the RE-induced thermal loading across the inner wall.

\section{Conclusions and Future work}\label{sec5}

This paper introduces the FIREWALL (Fast Integrated Runaway Electron WALL loads) model, a computationally efficient surrogate model for the global assessment of plasma-facing component heating by runaway electrons. FIREWALL bridges the gap between global RE dynamics simulations and computationally intensive high-fidelity wall damage modeling. Combining a database of \textsc{Geant4} energy deposition profiles with a one-dimensional heat-diffusion solver applied independently to each wall element, the model retains the essential dependence of volumetric heating on the RE energy and impact angle, while enabling the thermal response of detailed three-dimensional wall geometries to be quickly evaluated.

A comparison with a high-fidelity \textsc{Geant4}–MEMENTO thermal workflow \cite{Ratynskaia1} demonstrates that FIREWALL captures the energy-deposition and temperature profiles with sufficient accuracy to reliably identify the most critical wall regions and predict the onset of melting. Application to full-domain ITER RE termination simulations \cite{Bergstrom2024, Bandaru_2024} demonstrates the power of FIREWALL: rapid identification of where the first wall is threatened by melting and when during the RE loading this threshold is reached. Regions remaining below the melting point can be assessed directly, whereas regions reaching melting early in the loading are automatically identified as requiring subsequent high-fidelity treatment of melting, vaporization and the detailed component geometry. 

The present implementation is designed to be readily extended. Expansion of the \textsc{Geant4} database to additional plasma-facing materials, magnetic-field configurations and operating conditions will broaden its applicability beyond the tungsten ITER cases considered here. Moreover, the \textsc{Geant4} database is being extended to include the energy and angular distributions of backscattered secondary-particle populations generated by RE–wall interactions, providing input for future studies of their effects on RE formation and dynamics, as discussed in Ref. \cite{Ratynskaia2}. FIREWALL can also be coupled to RE-dynamics tools other than JOREK, including DREAM \cite{dream} and M3D-C1 \cite{M3d-c1,M3d-c1_re}, provided that the required wall-impact information is available. Finally, coupling to synthetic thermography and to photonic ray tracing \cite{Aumenier_2022, Aumeunier_2024} will enable more direct comparison between predicted wall heating and experimental observations.

\begin{acknowledgments}

\noindent Part of this work has been carried out within the framework of the EUROfusion Consortium, funded by the European Union via the Euratom Research and Training Programme (Grant Agreement No 101052200– EUROfusion). The views and opinions expressed are however those of the author(s) only and do not necessarily reflect those of the European Union or the European Commission. Neither the European Union nor the European Commission can be held responsible for them. SR acknowledges the financial support of the Swedish Research Council under Grant No.2025-05867. The computations and simulations were partly done on computing systems hosted by MPCDF and CINECA. The simulations were partly enabled by resources provided by the National Academic Infrastructure for Supercomputing in Sweden (NAISS) at the NSC (Link\"oping University) partially funded by the Swedish Research Council through grant agreement No\,2022-06725. This work explores the physics processes during plasma operation of the tokamak when disruptions take place; nevertheless the nuclear operator is not constrained by the results presented here. The views and opinions expressed herein do not necessarily reflect those of the ITER Organization.

\end{acknowledgments}

\section*{Author Declarations}

\subsection*{Conflicts of interest}

\noindent The authors have no conflicts to disclose.

\subsection*{Author contributions}

\noindent \textbf{Victor Svensson:} Conceptualization (lead); Data curation (lead); Formal analysis (lead); Investigation (lead); Methodology (lead); Software (lead); Validation (lead); Visualization (lead); Writing– original draft (lead); Writing– review $\&$ editing (supporting). \textbf{Tommaso Rizzi:} Conceptualization (supporting); Data curation (supporting); Formal analysis (supporting);  Methodology (supporting); Software (supporting); Validation (supporting); Visualization (supporting); Writing– original draft (supporting); Writing– review $\&$ editing (supporting). \textbf{Svetlana Ratynskaia:} Conceptualization (equal); Formal analysis (supporting); Investigation (supporting); Methodology (supporting); Validation (supporting); Writing– original draft (supporting); Writing– review $\&$ editing (lead). \textbf{Hannes Bergström:} Conceptualization (supporting); Data curation (supporting); Formal analysis (supporting); Investigation (supporting); Methodology (supporting); Validation (supporting); Visualization (supporting); Writing– review $\&$ editing (supporting). \textbf{Luca Greco:} Conceptualization (supporting); Software (supporting); \textbf{Matthias Hoelzl:} Conceptualization (equal); Investigation (supporting); Methodology (supporting); Writing– review $\&$ editing (supporting). \textbf{Panagiotis Tolias:}  Conceptualization (equal); Methodology (supporting); Writing– review $\&$ editing (supporting).

\section*{Data Availability Statement}

\noindent The code is available as an open source software under the LGPL-3.0 license at \url{https://github.com/iterorganization/FIREWALL}. FIREWALL is meant for broad use within the disruption and runaway electron modelling community. Data presented in this article is available from the authors upon reasonable request.

\section*{References}

%\nocite{*}
\bibliography{biblio}% Produces the bibliography via BibTeX.

\end{document}